\documentclass[prb,twocolumn,aps,showpacs]{revtex4}

\usepackage{graphicx}	
\usepackage{bm}	
\usepackage{amsmath,amssymb}
\usepackage{hyperref}
\usepackage{fix-cm,type1cm} 

\hypersetup{%
  breaklinks = {true},
  citecolor = {blue},
  colorlinks = {true},
  linkcolor = {red},
  pdfauthor = {\textcopyright\ The authors},
  pdfcreator = {\LaTeX\ and \flqq hyperref\frqq},
  pdffitwindow = {true},
  pdfmenubar = {true},
  pdfpagelayout = {SinglePage},
  pdfstartview = {Fit},
  pdftoolbar = {true},
  plainpages = {false},
}

\newcommand{\Hca}{\mathcal{H}}
\newcommand{\tp}{t_{\perp}}

\newcommand{\ra}{\rangle}

\DeclareMathOperator{\sign}{sign}

\begin{document}
\title{Bound states and magnetic field-induced valley splitting in gate-tunable \\ graphene quantum dots}

\author{Patrik Recher}
\affiliation{Instituut-Lorentz, Universiteit Leiden, P.O. Box 9506, 2300 RA Leiden, The Netherlands}

\author{Johan Nilsson}
\affiliation{Instituut-Lorentz, Universiteit Leiden, P.O. Box 9506, 2300 RA Leiden, The Netherlands}

\author{Guido Burkard}
\affiliation{Department of Physics, University of Konstanz, 78457 Konstanz, Germany}

\author{Bj{\"o}rn Trauzettel}
\affiliation{Institute for Theoretical Physics and Astrophysics, University of
W{\"u}rzburg, 97074 W{\"u}rzburg, Germany}

\date{October 2, 2008}
\begin{abstract} The magnetic field dependence of energy levels in
  gapped single- and bilayer graphene quantum dots (QDs) defined by
  electrostatic gates is studied analytically in terms of the Dirac
  equation. Due to the absence of sharp edges in these types of QDs,
  the valley degree of freedom is a good quantum number.
We show that its degeneracy is efficiently and controllably broken by
a magnetic field applied perpendicular to the graphene plane. This
opens up a feasible route to create well-defined and
well controlled spin- and valley-qubits in graphene QDs. We also point out the similarities and differences in the spectrum between single- and bilayer graphene quantum dots. Striking in the case of bilayer graphene is the anomalous bulk Landau level (LL) that crosses the gap which results in crossings of QD states with this bulk LL at large magnetic fields in stark contrast to the single-layer case where this LL is absent. The tunability of the gap in the bilayer case allows us to observe different regimes of level spacings directly related to the formation of a pronounced ``Mexican hat'' in the bulk bandstructure. We discuss the applicability of such QDs to control and measure the valley isospin and their potential use for hosting and controlling spin qubits.
\end{abstract}
\narrowtext \maketitle

\section{Introduction}

Graphene is one of the most promising materials for future
nano-electronics. \cite{geim_review,Castro_Review} This is related to
its truly two-dimensional character yielding perfect electron
confinement in one spatial dimension. In order to build functional
nano-devices such as single-electron transistors, quantum point
contacts, and quantum dots (QDs), additional confinement in the
remaining two spatial dimensions is needed.
Due to the absence of a gap in the spectrum, this is a rather
demanding task in both single- and bilayer graphene, in contrast
to electrostatically defined QDs
in semiconductors such as GaAs. One possibility of overcoming this
difficulty consists in etching or scratching nanostructures into graphene flakes. This has been done to experimentally study, for instance, transport through graphene nanoribbons, \cite{chen2007,Han2007,Li2008} single-electron transistors, \cite{Stampfer2008,Ponom2008} and, very recently, even QDs showing pronounced signatures of excited states. \cite{Stampfer2008b} Nevertheless, to increase the functionality of graphene nano-devices it is desirable to develop gate-tunable structures.

In this article, we study the energy spectrum of gate-tunable QDs both
in single-layer and bilayer graphene. In single-layer graphene, we
assume a constant gap in the whole system that might be introduced by
the underlying substrate. \cite{Brink2007,Lanzara_gap_2007} In bilayer
graphene, it is well-known that a gap can be generated by applying
different electrostatic potentials to the upper and lower layer,
\cite{McCann2006a,MacDonald_bilayergap_2006} which has already been
experimentally observed. \cite{Ohta06,Castro_PRL_2007,OHLMV07}
Once
there is a physical mechanism that gives rise to a gap, bound states exist in the presence of an electrostatic confinement potential. We focus on the magnetic field dependence of bound states in circularly symmetric QDs. Whereas previous work has analyzed bound states in single-layer graphene subjected to spatially inhomogeneous magnetic fields, \cite{Egger} we analytically study the magnetic-field dependence of bound states due to electrostatic (i.e. non-magnetic) confinement. A complementary numerical analysis has been done to study the Fock-Darwin spectrum of parabolic QDs in single-layer graphene \cite{Chakraborty} where only quasi-bound states but not true bound states exist. \cite{Silvestrov}

Most remarkably, we show how the valley degeneracy can be lifted by an external magnetic field applied perpendicular to the surface. This is of particular importance to form valley-filters, -valves,\cite{Rycerz2007} or -qubits,\cite{Recher2007}  and spin qubits \cite{Trauzettel2007} in graphene. To do so, it is essential to have full control both over spin and valley degrees of freedom and we show that a magnetic field is all that is needed to achieve this goal. Some of us have demonstrated that such a control can also be achieved in single-layer graphene ring structures with an Aharonov-Bohm flux applied. \cite{Recher2007} Here, the emphasis is on the more feasible situation of a constant magnetic field applied to the whole system. We would also like to mention that the broken valley degeneracy has an interpretation in terms of a magnetic moment that depends on the valley isospin.\cite{Niu_ValleyPrl_2007} Our models for single- and bilayer graphene QDs are appropriate for the physical situation of a smooth crossover between the dot region and the barrier. Therefore, atomically sharp edges do not play any role in our analysis which seems to be the most relevant case for possible experimental realizations of gate-tunable QDs in graphene.  Indeed, in Ref.~\onlinecite{Ponom2008} the absence of a four-fold level degeneracy in the transport data---due to valley and spin degrees of freedom---was attributed to inter-valley scattering at the atomically sharp edges whereas the absence of spin degeneracy could result from spin scattering at dangling bonds at the edge of the QD. Both possible sources for an {\it uncontrolled} lifting of degeneracies are not relevant for our QD realizations.

In the regime of strong magnetic fields, the bound states of single- and bilayer graphene merge into the appropriate bulk Landau levels (LLs) as expected. However, the nature of these LLs are quite different for the cases of single layer graphene versus bilayer graphene. In the bilayer case, one of the LLs crosses the gap with increasing magnetic field. QD bound states can cross this LL at large magnetic fields which leads to certain constraints to form operational QDs.

The paper is organized as follows. In Sec.~II, we discuss our model
and the results for single-layer QDs. In Sec.~III, we treat the
bilayer case, in Sec. IV we discuss possible applications of
our results for the emerging field of valleytronics and to spin-based
qubits in graphene, and, in Sec.~V, we draw our conclusions.

\section{QD in single layer graphene}
\begin{figure}[h]
\vspace{0.4cm}
\begin{center}
\includegraphics[width=0.6\columnwidth]{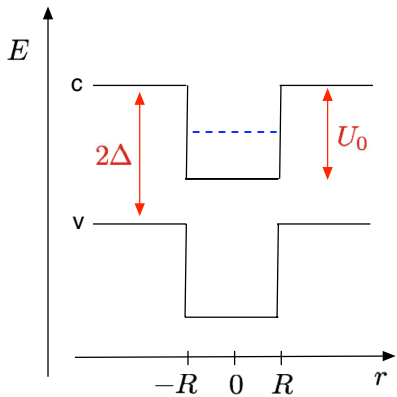}
\end{center}
\vspace{-0.4cm}
\caption{QD in single-layer graphene with a constant mass term $\Delta$. An electrostatic potential with height $U_{0}$ gives rise to bound states (dashed line) in the conduction band (c) defining a QD of radius $R$. Note that the confining potential $U(r)$ is repulsive for holes in the valence band (v).}
\label{QD1}
\end{figure}
In this section we study graphene in the presence of a constant mass term $\Delta$ (inducing a gap $2\Delta$) that might be introduced by the underlying substrate. \cite{Brink2007,Lanzara_gap_2007} The QD is defined by gates introducing an electrostatic confining potential for electrons in the conduction band (see Fig.~\ref{QD1}). We also include a homogeneous magnetic field $B$ perpendicular to the graphene plane.

The Hamiltonian in the valley-isotropic
form is given by \cite{Carlographenereview}
\begin{equation} \label{htau}
H_\tau = H_0 + \tau \Delta \sigma_z + U(x,y) ,
\end{equation}
where $H_0 = v({\bm p} + e{\bm A}) \cdot {\bm \sigma}$, ${\bm B} =
\nabla \times { \bm A} = (0,0,B)$, $v=10^6$ m/s is the Fermi velocity and $\tau = \pm$ differentiates the
two valleys $K$ and $K'$. We choose the symmetric gauge ${\bm A} =\frac{B}{2} (-y,x,0)$ and assume a circular symmetry in the
confinement potential $U(x,y)=U(r)$ with $r=\sqrt{x^2 + y^2}$. The vector operator ${\bm \sigma}$ acts on the $A,B$ sublattice components of the spinor wave function and its vector components are given by the standard Pauli matrices.

$H_0$ may be transformed into polar coordinates [$(x,y)=(r
\cos\varphi,r \sin\varphi)$] (with $\hbar=1$)
\begin{multline}
H_0 = -i v \left( \begin{array}{cc} 0 & e^{-i \varphi} \\ e^{i
\varphi} & 0 \end{array} \right) \partial_r \\+ v \left(
\begin{array}{cc} 0 & - e^{-i \varphi} \\ e^{i \varphi} & 0
\end{array} \right) \left( \frac{1}{r} \partial_\varphi +
\frac{ieBr}{2} \right) .
\end{multline}
Since $H_\tau$ commutes with the total angular momentum operator $J_{z}=-i\partial_{\varphi}+\sigma_{z}/2$, the energy eigenspinors can be chosen to be eigenstates of $J_{z}$
\begin{equation}
\Psi^\tau(r,\varphi) = e^{i(j-1/2) \varphi} \left( \begin{array}{c}
\chi_A^\tau(r) \\ \chi_B^\tau (r) e^{i \varphi} \end{array}
\right) ,
\end{equation}
with $j$ the eigenvalue of $J_{z}$ which has to be an half-odd integer.
\subsection{Bound state solutions}
To solve the eigenvalue problem $H_\tau \Psi^\tau(r,\varphi) = E
\Psi^\tau(r,\varphi)$ we have to analyze
\begin{equation}
\label{H1}
\tilde{H}_\tau(r) \chi^\tau(r) = E \chi^\tau(r),
\end{equation}
with $\chi^\tau(r)=(\chi_A^\tau(r),\chi_B^\tau(r))^{T}$ and
\begin{multline}
\tilde{H}_\tau (r) = -iv \sigma_x \partial_r + \tau \Delta
\sigma_z + U(r) + \\v \sigma_y  \left(
\begin{array}{cc} \frac{j-1/2}{r} + \frac{eBr}{2} & 0 \\ 0 &
\frac{j+1/2}{r} + \frac{eBr}{2} \end{array} \right) .
\end{multline}
First, we solve Eq.~(\ref{H1}) with a constant $U(r)=U_{0}$. Defining $\epsilon\equiv E-U_{0}$ and $b\equiv eB/2$,
we obtain the following decoupled second order
differential equations
\begin{equation} \label{secdif}
r^2 \partial_r^2 \chi^\tau_\sigma(r) + r \partial_r
\chi^\tau_\sigma(r) = (b^2 r^4 + a_\sigma r^2 + n_\sigma^2)
\chi^\tau_\sigma(r),
\end{equation}
with $\sigma = \pm 1$, the upper sign corresponding to the $A$, the
lower to the $B$ sublattice, the coefficients entering
Eq.~(\ref{secdif}) can be expressed as
\begin{eqnarray}
a_\sigma &=&  2b (j+\sigma/2) - (\epsilon^2 - \Delta^2)/v^2 ,\\
n_\sigma &=&  |j-\sigma/2| .
\end{eqnarray}
%
%
Note that Eq.~(\ref{secdif}) does not depend on the valley index $\tau$ anymore. However,  $\chi^\tau_\sigma(r)$ depends on $\tau$ through Eq.~(\ref{H1}). The solutions to
Eq.~(\ref{secdif}) are the confluent hypergeometric functions
$M(a,b,z)$ and $U(a,b,z)$. The boundstate solutions for the QD have the form
\begin{multline}
\label{ansatz}
\chi_\sigma^\tau(r) = 2^{(1+n_\sigma)/2} e^{-br^2/2} r^{n_\sigma}\\
\times \left\{ \begin{array}{cc} \alpha_\sigma U(q_\sigma,
1+n_\sigma,br^2) & , \; r>R \; ,
\\  \beta_\sigma M(q_\sigma,
1+n_\sigma,br^2) & , \; r<R \; , \end{array} \right.
\end{multline}
where $q_\sigma \equiv \frac{1}{4} [\frac{a_\sigma}{b} + 2
(1+n_\sigma)]$. Eq.~(\ref{ansatz}) is the general solution for waves that are regular at the origin and which decay exponentially as $r\rightarrow \infty$. We want to find the bound states for the following
hard-wall potential
\begin{equation}
\label{eq:Uform}
U(r) =  \left\{ \begin{array}{cc} U_0 & , \; r>R \; ,
\\ 0 & , \; r<R \; ,
\end{array} \right.
\end{equation}
and define the corresponding energies as $\epsilon_< \equiv E$ and
$\epsilon_> \equiv E- U_0$. \\
The ratios $\alpha_{B}/\alpha_A$ and $\beta_B/\beta_A$ in  Eq.~(\ref{ansatz}) are fixed by the coupled first-order
differential equation Eq.~(\ref{H1}).
This provides us with the general solutions for $r<R$ and $r>R$.
The matching conditions of the spinors at $r=R$ gives then the eigenvalues and eigenfunctions of the bound states.

For $j>0$, we obtain the following characteristic equation for the allowed eigenenergies $E$ of the QD
\begin{widetext}
\begin{equation}
\label{eq:E1}
\xi_{>}^{+}\,M(q_<,j+1/2,x)\,U(q_{>},j+3/2,x)-\xi_{<}^{+}\,M(q_{<},j+3/2,x)\,U(q_{>},j+1/2,x)=0,
\end{equation}
and for $j<0$ we obtain
\begin{equation}
\label{eq:E2}
\xi_{>}^{-}\,
M(q_<,-j+3/2,x)\,U(q_{>}-1,-j+1/2,x)-\xi_{<}^{-}\,M(q_{<}-1,-j+1/2,x)\,U(q_{>},-j+3/2,x)=0,
\end{equation}
\end{widetext}
where $x\equiv bR^2$=$(1/2)(R/l_{B})^2$ with $l_{B}=\sqrt{\hbar/eB}$ the magnetic length. Without loss of generality, we choose $B$ positive. The bound state levels for $B$ negative can be obtained from the symmetry ${\tilde H}_{\tau}(j,B)={\tilde H}_{-\tau}(-j,-B)$. We further introduced the parameters $q_{<,>}=(j-1/2)\,\theta(j)+1-(\epsilon_{<,>}^2-\Delta^2)/4bv^2$, $\xi_{<}^{+}=(\epsilon_{<}-\tau\Delta)/4(j+1/2)$, $\xi_{>}^{+}=b/(\epsilon_{>}+\tau\Delta)$, $\xi^{-}_{<}=(j-1/2)/(\epsilon_{<}+\tau\Delta)$ and $\xi_{>}^{-}=1/(\epsilon_{>}+\tau\Delta)$ with $\theta(x)$ the Heaviside function.

In the limit of small magnetic fields $(x\ll 1)$, the hypergeometric functions reduce to Bessel functions (see Ch. 13 in Ref.~\onlinecite{AS1965})
\begin{multline}
\label{E5}
M(q_{<},n,x)=\Gamma(n)\left(-x\, q_{<}\right)^{(1-n)/2} \\\times J_{n-1}\left(2\sqrt{-x\, q_{<}}\right),
\end{multline}
and
\begin{multline}
\label{E6}
U(q_{>},n,x)=\frac{2}{\Gamma(1+q_{>}-n)}\left(x\, q_{>}\right)^{(1-n)/2} \\\times K_{n-1}\left(2\sqrt{x\, q_{>}}\right),
\end{multline}
where we have introduced the QD level spacing $\delta=\hbar v/R$.
For $B=0$, the characteristic equation for $j > 0$ (Eq.~(\ref{eq:E1})) becomes
\begin{multline}
\label{E7}
\frac{\epsilon_{<}-\tau\Delta}{\epsilon_{>}-\tau\Delta}\sqrt{\frac{\Delta^2-\epsilon_{>}^2}{\epsilon_{<}^2-\Delta^2}}\\\times J_{j+1/2}\left(\frac{2}{\delta}\sqrt{\epsilon_{<}^2-\Delta^2}\right)K_{j-1/2}\left(\frac{2}{\delta}\sqrt{\Delta^2-\epsilon_{>}^2}\right)\\
+J_{j-1/2}\left(\frac{2}{\delta}\sqrt{\epsilon_{<}^2-\Delta^2}\right) K_{j+1/2}\left(\frac{2}{\delta}\sqrt{\Delta^2-\epsilon_{>}^2}\right)=0.
\end{multline}
For $j<0$ (Eq. (\ref{eq:E2})), we obtain Eq.~(\ref{E7}) with $j\rightarrow -j$ and $\tau\rightarrow -\tau$.
\begin{figure}[h]
\vspace{0.4cm}
\begin{center}
\includegraphics[width=0.8\columnwidth]{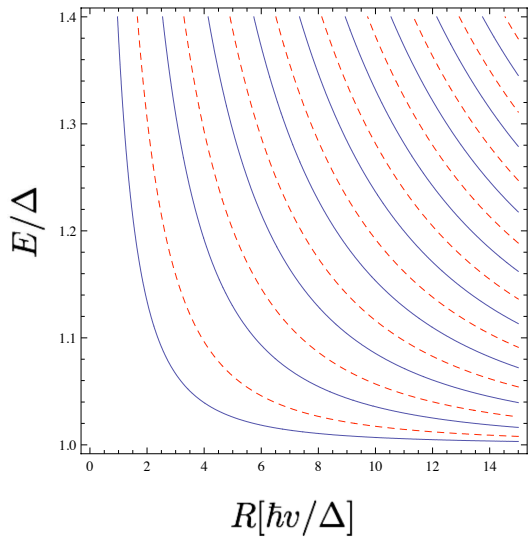}
\end{center}
\vspace{-0.4cm}
\caption{Bound state levels as a function of the QD radius $R$ with $U_{0}=\Delta$ and for $j=1/2$ at zero magnetic field. Full lines correspond to $\tau=+1$, dashed lines correspond to $\tau=-1$.}
\label{zerofield}
\end{figure}

Even in the limit of zero magnetic field, the characteristic equation (Eq~(\ref{E7})) cannot be solved in closed form in general. However,
the fact that $E(j,\tau)\neq E(-j,\tau)$, but $E(j,\tau)= E(-j,-\tau)$ lies at the heart of our current approach to control the valley degeneracy by a magnetic field.
The first statement is a consequence of effective time-reversal symmetry (eTRS) breaking within a single valley by a finite mass $\Delta$. \cite{Berry1987}
Formally, $[H_{\tau},{\tilde T}]\neq 0$ where $\tilde{T}=i\sigma_{y}{\cal C}$ with ${\cal C}$ the operator of complex conjugation.
The second statement is that the true TRS (which couples the two valleys) is not broken by a boundary alone, i.e. at $B=0$ (see also subsection IV A).
\begin{figure}[h]
\label{singlelayerfig1}
\vspace{0.4cm}
\begin{center}
\includegraphics[width=0.8\columnwidth]{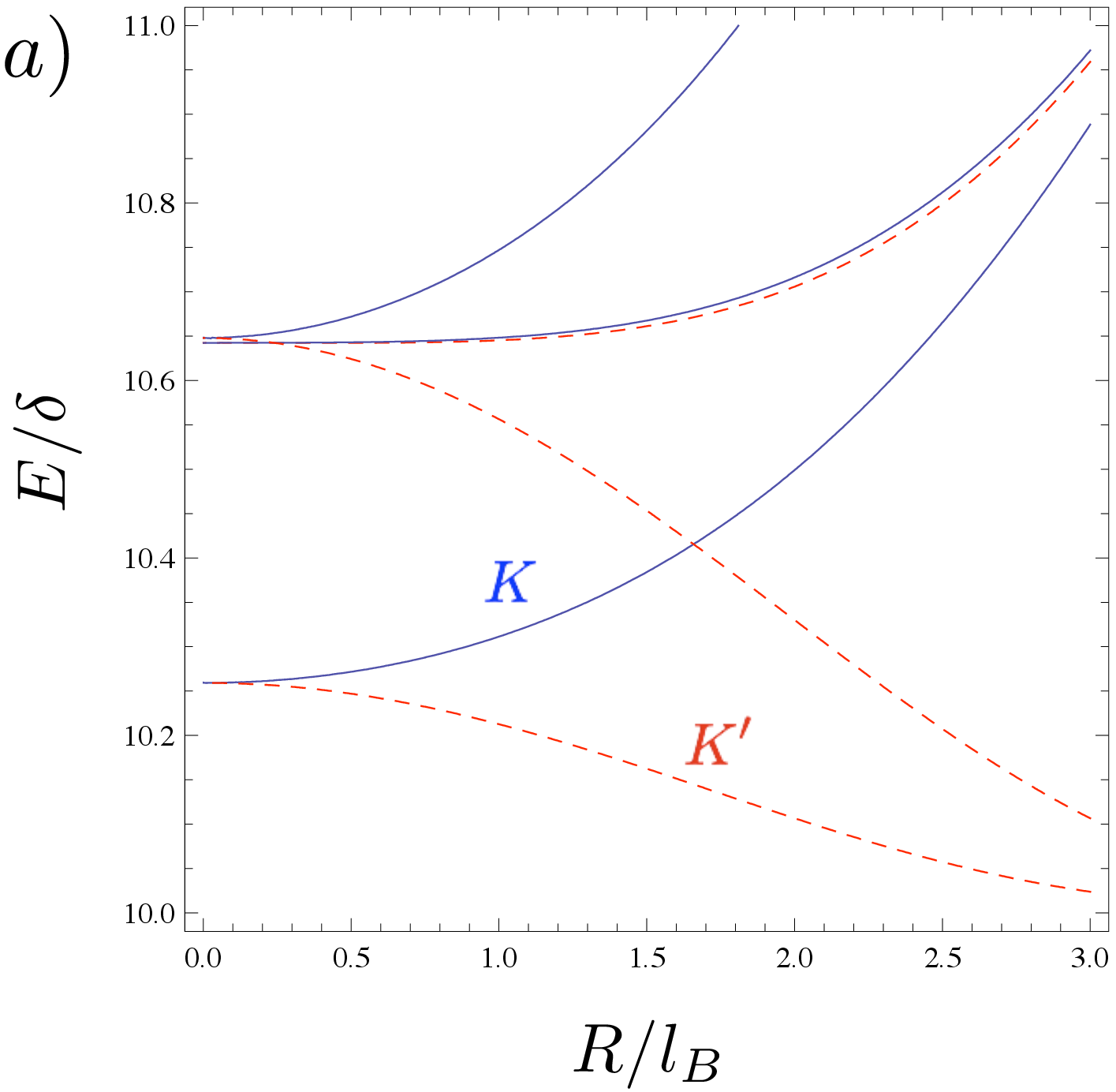}
\end{center}
\begin{center}
\includegraphics[width=0.8\columnwidth]{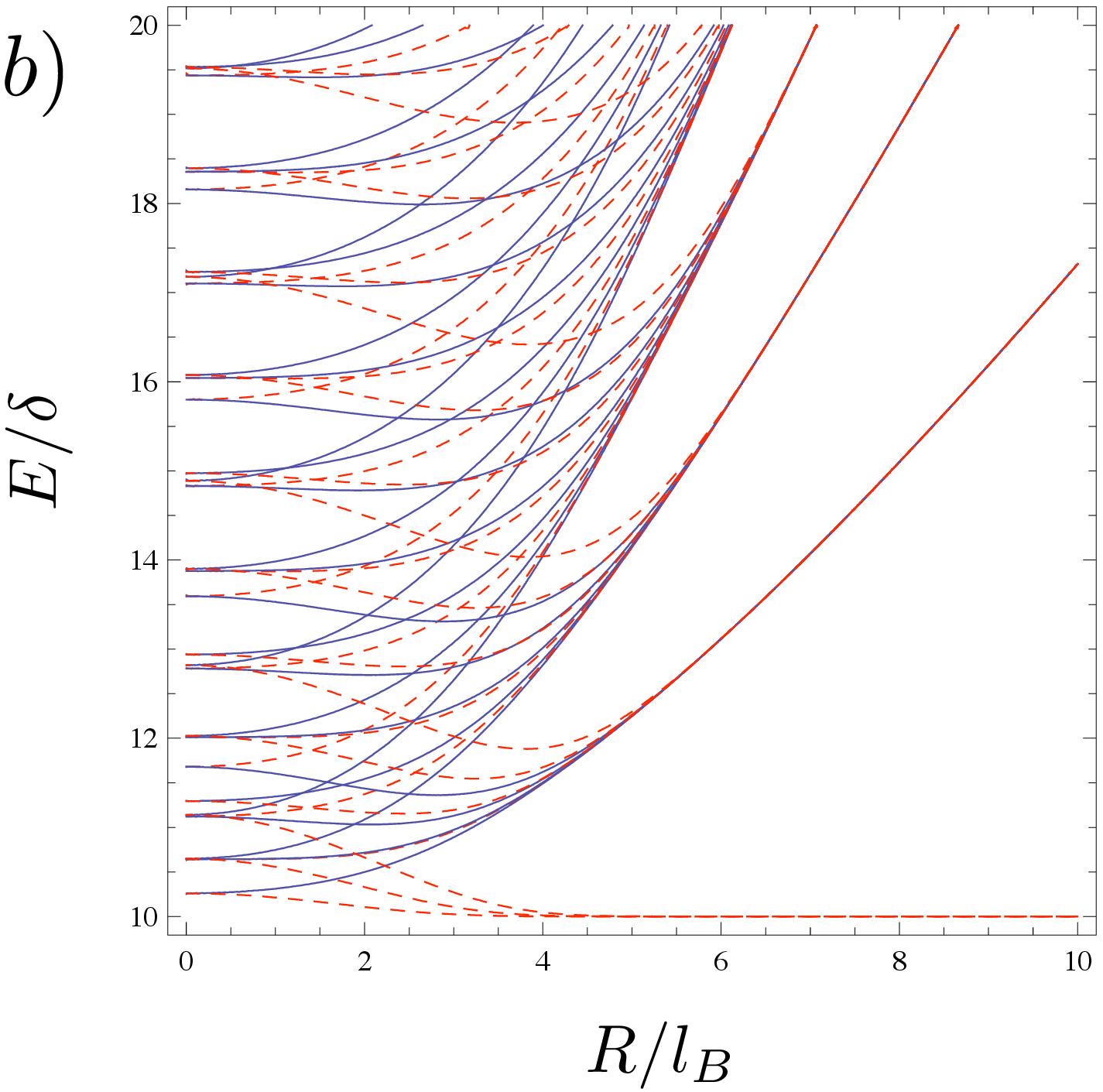}
\end{center}
\vspace{-0.4cm}
\caption{a) Numerical evaluation of characteristic equations (\ref{eq:E1}) and (\ref{eq:E2}) as a function of $R/l_{B}$ with $l_{B}=(\hbar/eB)^{1/2}$ the magnetic length and $R$ the QD radius. We use $\Delta=10\,\delta$ and $U_{0}=\Delta$. a) The parameter regime of small $B$-fields where we observe a breaking of the level degeneracy. The full lines are for $\tau=1$ and
dashed lines are for $\tau=-1$ corresponding to the two valleys of graphene. b) Same parameters as in a),  but for larger magnetic fields. The energy levels converge to the bulk Landau levels with increasing $R/l_{B}$. }
\label{singlelayerfig}
\end{figure}

\subsection{Results}

We first consider zero magnetic field. In Fig.~\ref{zerofield} we show
the energy levels of the QD as a function of the dot radius $R$,
evaluating  Eq.~(\ref{E7}) for $j=1/2$. Full lines and dashed lines
correspond to the two valleys. Due to the symmetry $E(j,\tau)=
E(-j,-\tau)$, the two set of curves display also the cases $j=1/2$ and
$j=-1/2$ in the same valley. The different solutions for the dashed
and full lines are therefore a direct consequence of eTRS breaking in a
single valley at zero magnetic field.
However, if both signs of $j$ were included, one would observe that the valley
degeneracy was not broken at $B=0$.

In Fig.~\ref{singlelayerfig} we show the bound states of the QD as a function of magnetic field evaluating the characteristic equations Eqs.~(\ref{eq:E1}) and (\ref{eq:E2}) numerically.
In Fig.~\ref{singlelayerfig}(a) we show the low-lying bound states in the conduction band. Note that the valley-degeneracy (or orbital degeneracy) is broken at finite magnetic field. The largest level spacing between the (non-degenerate) groundstate and first excited state we estimate from Fig.~\ref{singlelayerfig}(a) to be at $R/l_{B}\sim 1.8$ and is about 165 meV/$R$[nm] for the parameters used in Fig.~\ref{singlelayerfig}. At $R/l_{B}\sim 1.8$ we obtain for the Zeeman splitting $\Delta_z=g\mu_{B}B\sim 200$ meV/$R^2$[nm] using $g=2$ which shows that {\it the level spacing is always larger than the Zeeman energy for reasonable dot sizes}.

Considering a QD with $R=25$ nm, we obtain a valley splitting $\Delta_{K,K'}$ at $R/l_{B}\sim 1.8$ of about 6.6 meV corresponding to 77 K, being much larger than $4$ K, the temperature achieved by cooling with liquid helium. The necessary magnetic field corresponding to $R/l_{B}= 1.8$ is $B$=3.41 T (and $B=0.85$ T for $R=50$ nm with $\Delta_{K,K'}\sim 3.3$ meV)  which is also easily achievable in the laboratory. A gap of size 0.23 eV has been concluded from ARPES data in graphene on top of a SiC substrate. \cite{Lanzara_gap_2007} Therefore, the gap $\Delta$ and also the confining potential step height $U_{0}$ could easily be larger than the QD level spacing $\delta$ which is about 26 meV.
These results suggest that such QDs confined in graphene would be an ideal host for spin qubits where the orbital degeneracy is controllable by a magnetic field.

In Fig.~\ref{singlelayerfig}(b) we show the merging of the QD states with the bulk Landau levels (LLs)
\begin{equation}
E_{n}=\pm\delta\sqrt{(\Delta/\delta)^2+2n(R/l_{B})^2};\,\,n=1,2,3,...
\end{equation}
with increasing magnetic field. Note in particular, that there is a zero mode LL at $E=-\tau\Delta$ which lies entirely in one valley. \cite{Haldane1988}

In the next section we consider QDs in bilayer graphene where a voltage tunable mass gap is possible.

\section{QD in bilayer graphene}

\begin{figure}[h]
\vspace{0.4cm}
\begin{center}
\includegraphics[width=0.9\columnwidth]{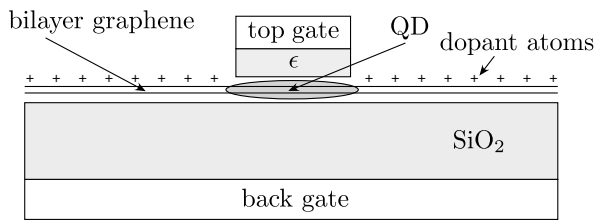}
\end{center}
\vspace{-0.4cm}
\caption{QD in bilayer graphene: A back gate and dopants on top of the bilayer control the voltage $V$ between the layers---leading to a controllable gap opening---as well as the Fermi energy (band filling). An additional top gate allows to induce a spatially inhomogeneous electrostatic potential $U(r)$ analogous to the single-layer model which leads to bound states in the conduction band (or valence band) of the bilayer. Another possibility is to use a split top gate (instead of a combination of top gate and dopants) to achieve a similar confinement.}
\label{fig:QD2}
\end{figure}

In bilayer graphene, an electric field perpendicular to the layers generates a gap in the spectrum in a similar fashion to the staggered sublattice potential in the single layer. In this section, we will investigate the bilayer analogue of the single layer QD studied in the previous section, as shown in Fig.~\ref{fig:QD2}.
We use the simplest nontrivial form of the Hamiltonian that captures the most important features of the spectrum and calculate the quantized energy levels of the QD as a function of the magnetic field and the relevant parameters of the band structure and the QD. The approximate Hamiltonian (we use) correctly describes the crucial formation of an electronic gap in biased bilayer graphene.\cite{Nilsson_long_2008} We briefly discuss the issue of neglected terms in Sec.~\ref{rbsl}.

\subsection{Solving for the energy levels}

We separate the Hamiltonian in the bilayer into two parts: $\Hca = \Hca_0 + \Hca_1^{\tau}$. $\Hca_0$ encodes the motion of the electrons within the planes and is given by two copies of the Dirac equation. In the valley-isotropic representation it takes on the form ($\hbar=v=1$)
\begin{equation}
\label{eq:H0b_1}
\Hca_0 =
\begin{pmatrix}
0 & p_x + i p_y & 0 & 0 \\
p_x - i p_y & 0 & 0 & 0 \\
0 & 0 &  0 & p_x - i p_y  \\
0 & 0 & p_x + i p_y & 0
\end{pmatrix},
\end{equation}
in both valleys. Like in the case of the single layer we add a magnetic field by the minimal coupling prescription ${\bm p}\rightarrow( {\bm p}+e{\bm A})$ with ${\bm A}=(B/2)(-y,x,0)$.
The other part of the Hamiltonian (i.e. $\Hca_1^{\tau}$) encodes the biasing field and
the hopping $\tp$ between the two planes. The interplane hopping matrix element $\tp$ has recently been measured to be $\tp = 0.40$ eV. \cite{Li_2008_bilayer,Zhang_2008_bilayer} In the simplest approximation
we may take
\begin{equation}
\label{eq:H1b_1}
\Hca_1^{\tau} =
\begin{pmatrix}
\frac{\tau V}{2} & 0 & \tp & 0 \\
0 & \frac{\tau V }{2} & 0 & 0 \\
\tp & 0 &  -\frac{ \tau V}{2} & 0  \\
0 & 0 & 0 & -\frac{\tau V}{2}
\end{pmatrix} + U(r) {\bm 1},
\end{equation}
with $U(r)$ the applied electrostatic potential profile again given by Eq.~\eqref{eq:Uform}. The index $\tau = \pm 1$ again distinguishes the two valleys (note that in the valley-isotropic representation the basis is chosen such that the two planes in the bilayer are exchanged in the spinors that describes different valleys). In Ref.~\onlinecite{Peeters_dots_2007}, the same Hamiltonian ${\cal H}$ {\it at zero magnetic field} was used. However, the confinement described in Eq.~(\ref{eq:H1b_1}) by $U(r)$ was achieved in Ref.~\onlinecite{Peeters_dots_2007} by a position dependent "mass term" $V(r)$, instead. 

To diagonalize $\Hca$ (i.e. to find the eigenspinors $\Psi$ that fulfill $\Hca \Psi = E \Psi$) we go to cylindrical coordinates in which the states are easily classified according to their conserved value of total angular momentum $m$ ($m$ being an integer). More explicitly, we factor out the angular dependence of the states according to
\begin{equation}
\label{eq:transform1}
\Psi =
\frac{e^{ i m \varphi}}{\sqrt{r}}
\begin{pmatrix}
1 & 0 & 0 & 0 \\
0& e^{- i \varphi} & 0 & 0 \\
0& 0 & 1 & 0 \\
0 & 0 & 0 & e^{ i \varphi}
\end{pmatrix}
\Psi_1.
\end{equation} 
Note that the angular momentum in the bilayer case is an integer $m$, in contrast to the half-odd integer $j$ in the single layer case, which reflects the different pseudospins in the bilayer (pseudospin 1) and single-layer (pseudospin 1/2).
With the definitions $j = m+1/2$ and $s=\sign(B)$, the Hamiltonian $\Hca_0$, which now acts on  $\Psi_1$, can be written as
\begin{widetext}
\begin{equation}
\label{eq:H0b_trans1}
\Hca_0 =
\frac{1}{i \sqrt{2} l_B }
\begin{pmatrix}
0 & \partial_{\xi} - (j-1)/ \xi -s \xi & 0 & 0 \\
\partial_{\xi} +(j-1)/ \xi +s \xi & 0 & 0 & 0 \\
0 & 0 &  0 & \partial_{\xi} + j / \xi +s \xi  \\
0 & 0 & \partial_{\xi} - j / \xi -s \xi & 0
\end{pmatrix}.
\end{equation}
\end{widetext}
In the latter equation, we have introduced the dimensionless coordinate $\xi = r /(\sqrt{2} l_B)$, where $l_B = \sqrt{\hbar / (e |B|)}$ is the magnetic length.
The eigenvalue problem can now be solved by using the general
solutions of the ordinary differential equation imposed by $\Hca_0$. The general solutions can be conveniently written in a simple way using the following functions (valid for all integers $m$ and $s=\pm 1$):
\begin{widetext}
\begin{equation} \label{soluti}
\phi_{m+\alpha}^{s}  \equiv  e^{-\xi^2 /2} \xi^{|m+\alpha|+1/2} M ( [ |m+\alpha|+1+s (m-1-\alpha) ]/2+\kappa^2/4 ,1+|m+\alpha|,\xi^2) /\Gamma(1+|m+\alpha|).
\end{equation}
\end{widetext}
These solutions are regular at the origin and are used for $r\leq R$. Note that $\kappa$ is an arbitrary parameter, which is chosen to be proportional to the energy eigenvalue of the first subblock of the matrix in Eq.~\eqref{eq:H0b_trans1}, i.e. $\Hca_0\Psi_1=-i\kappa \Psi_1/\sqrt{2}l_{B}$ for the first two components of $\Psi_1$. (This choice is motivated by mathematical convencience to simplify the recursion relations in Eqs.~(\ref{reca})---(\ref{recd}) below.)
In addition to Eq.~(\ref{soluti}), there are solutions that are irregular at the origin but vanish exponentially for $r\rightarrow \infty$ which we use for $r > R$. These solutions are given by the same expression as Eq.~(\ref{soluti}) with the substitution
\begin{multline}
M ( [ |m+\alpha|+1+s (m-1-\alpha) ]/2+ \\ \kappa^2/4 ,1+|m+\alpha|,\xi^2) /\Gamma(1+|m+\alpha|)
\\ \rightarrow U ( [|m+\alpha|+1+s (m-1-\alpha) ]/2+\kappa^2/4 ,1+|m+\alpha|,\xi^2).
\end{multline}
For both types of solutions one can show the following identities
by straightforward manipulations using the recursion relations
for the confluent hypergeometric functions (see e.g chapter 13 of Ref.~\onlinecite{AS1965}).
\begin{subequations}
\begin{eqnarray}
(\partial_{\xi} - (j-1)/ \xi -s \xi) \phi_{m-1}     &=&   a_{1}^{s} \,\phi_{m}, \label{reca}\\
(\partial_{\xi} +(j-1)/ \xi +s \xi) \phi_{m}  & =&  a_{2}^{s} \,\phi_{m-1}, \\
(\partial_{\xi} + j / \xi + s\xi) \phi_{m+1}  & =&  a_{3}^{s} \,\phi_{m}, \\
(\partial_{\xi} - j / \xi - s\xi) \phi_{m}  & =&  a_{4}^{s}\, \phi_{m+1}.
\label{recd}
\end{eqnarray}
\end{subequations}
For $r\leq R$ and for $m\geq 1$ we obtain
\begin{subequations}
\begin{eqnarray}
a_{1}^{s}   &=&   \kappa^2/2 ,\\
a_{2}  & =&  2  ,\\
a_{3}^{s}  & =&  2 ,\\
a_{4}^{s}  & =&  (\kappa^2-4s)/2.
\end{eqnarray}
\end{subequations}
For $r\leq R$ and $m=0$,
\begin{subequations}
\begin{eqnarray}
a_{1}^{s}    &=&   2 ,\\
a_{2}^{s}  & =& \kappa^2/2 ,\\
a_{3}^{s}  & =&  2  ,\\
a_{4}^{s} & =&  (\kappa^2-4s)/2,
\end{eqnarray}
\end{subequations}
and for $r\leq R$ and $m\leq -1$,
\begin{subequations}
\begin{eqnarray}
a_{1}^{s}     &=&   2  ,\\
a_{2}^{s}  & =&  \kappa^2/2 ,\\
a_{3}^{s} & =&  (\kappa^2-4s)/2 ,\\
a_{4}^{s}  & =&  2.
\end{eqnarray}
\end{subequations}
For $r>R$ and all integer $m$ we obtain
\begin{subequations}
\begin{eqnarray}
a_{1}^{s}     &=&   -[(s+1)+\kappa^2(1-s)/4]  , \\
a_{2}^{s}  & =&  -[(1-s)+\kappa^2(1+s)/4]  , \\
a_{3}^{s} & =&   -[(1-s)+(\kappa^2/4-1)(1+s)] , \\
a_{4}^{s}  & =&   -[(s+1)+(1+\kappa^2/4)(1-s)]  .
\end{eqnarray}
\end{subequations}
Therefore, by combining the solutions in the form
\begin{equation}
\label{eq:transform2}
\Psi_1 =
\begin{pmatrix}
\phi_{m} & 0 & 0 & 0 \\
0& \phi_{m-1} & 0 & 0 \\
0& 0 & \phi_{m} & 0 \\
0 & 0 & 0 & \phi_{m+1}
\end{pmatrix} \Psi_2 ,
\end{equation}
the $\Hca_0$ part of the Hamiltonian (now acting on $\Psi_2$) can be replaced by:
\begin{equation}
\Hca_0 =
\frac{1}{i \sqrt{2} l_B }
\begin{pmatrix}
0 & a_{1}^{s} & 0 & 0 \\
a_{2}^{s} & 0 & 0 & 0 \\
0 & 0 &  0 & a_{3}^{s}  \\
0 & 0 & a_{4}^{s} & 0
\end{pmatrix}.
\end{equation}
We now note that the transformations in Eq.~\eqref{eq:transform1}
and Eq.~\eqref{eq:transform2} commute with the part of the
Hamiltonian $\Hca_1^{\tau}$ of Eq.~\eqref{eq:H1b_1}. Therefore the task of finding the
eigenvectors is transformed into the simple problem of finding the eigenvectors of
a $4 \times 4$ matrix. Explicitly, the eigenvalue problem is equivalent to finding the
non-trivial solutions of
\begin{widetext}
\begin{equation}
\begin{pmatrix}
\frac{\tau V}{2}+U(r) - E & -i a_{1}^{s}/\sqrt{2}l_{B}& \tp & 0 \\
- i a_{2}^{s} /\sqrt{2}l_{B}& \frac{\tau V}{2}+U(r) - E & 0 & 0 \\
\tp & 0 &  -\frac{\tau V}{2} +U(r)-E & -  i a_{3}^{s}/\sqrt{2}l_{B}  \\
0 & 0 & -i a_{4}^{s}/\sqrt{2}l_{B} & -\frac{\tau V}{2} +U(r)- E
\end{pmatrix}
\Psi_2 = 0 .
\label{eq:4x4forPsi2}
\end{equation}
\end{widetext}
The non-trivial solutions are identified by finding the values of $\kappa^2$ such
that the determinant of the matrix is zero. Given the values of  $E$, $V$, $\tp$, and $B$
this amounts to solving a quadratic equation for $\kappa^2$ inside ($U(r)=0$) and outside ($U(r)=U_{0}$) the QD with the result
\begin{multline}
\label{kappas}
\frac{\kappa_{<,>}^2}{2l_{B}^2}=\frac{s}{l_{B}^2}-\epsilon_{<,>}^{2}-\frac{V^2}{4}\\
\pm\sqrt{\tp^{2}\left(\epsilon_{<,>}^{2}-\frac{V^2}{4}\right)+\left(\epsilon_{<,>}\tau V-\frac{s}{l_{B}^{2}}\right)^2},
\end{multline}
which is independent of $m$. The energies $\epsilon_{>}$ and $\epsilon_{<}$ are defined in Sec.~II.
With the knowledge of $\kappa^2$, we can easily find the corresponding eigenvector
$\Psi_2$. Finally we may use Eq.~\eqref{eq:transform1} and Eq.~\eqref{eq:transform2}  to recover the eigenvector $\Psi$ in the original basis. A similar procedure was used previously in Ref.~\onlinecite{Nilsson_long_2008} in the case of zero magnetic field.

Given the eigenvectors inside and outside the dot the bound state solutions of the full problem are those where the two pairs of solutions can be matched at the boundary of the dot. This is most easily tested by computing the determinant of the matrix built up by the four relevant eigenvectors evaluated at $r=R$, where $R$ is the radius of the dot. The zeros of the determinant as a function of the energy (inside of the gap at $r \rightarrow \infty$) determines the bound states and their energies. The condition of having the determinant equal to zero is the bilayer analogue of Eqs.~\eqref{eq:E1} and \eqref{eq:E2} for the single layer case and can straightforwardly be computed numerically although the analytic expression is long and cumbersome.

\subsection{Landau levels in biased graphene bilayer}

In this section we briefly review the properties of a biased graphene bilayer in a magnetic field. From the point of view of the QD there exists one level that is of particular importance since it crosses the gap with increasing magnetic field (see also Refs.~\onlinecite{PPV07,Castro_Long_2008}).
In the presence of a magnetic field, the Hamiltonian matrix in a homogenous system can be written as ($\tau=+1$)
\begin{equation}
\label{eq:H0_mangetic}
\Hca_0 =
\begin{pmatrix}
V/2 & \gamma a^{\dagger} & \tp & 0 \\
\gamma a & V/2 & 0 & 0 \\
\tp & 0 &  -V/2 & \gamma a \\
0 & 0 & \gamma a^{\dagger} & -V/2
\end{pmatrix},
\end{equation}
in the Landau gauge ${\bf A}=(0,Bx,0)$ and for a particular sign $(s=+1)$ of the magnetic field (changing the sign would just take $a \leftrightarrow a^{\dagger}$ and the same spectrum but with $V \rightarrow -V$ is obtained). Explicitly, $a^{\dagger}=is\sqrt{e|B|/2\hbar}\,[x-i(s/e|B|)(-i\hbar \partial_{x})]+ip_{y}$ with $p_y$ a c-number due to translational invariance in y-direction. We have defined $\gamma = v_F \hbar \sqrt{2}/l_B$. The eigenstates can then be formed by a spinor of the form
\begin{equation}
\Psi = [a_{A1} |n\ra , a_{B1} |n-1 \ra , a_{A2} |n\ra , a_{B2} |n+1\ra  ]^T.
\end{equation}
With this choice the operator matrix in Eq.~\eqref{eq:H0_mangetic} becomes a matrix of numbers acting on the spinor $\tilde{\Psi} = [a_{A1} , a_{B1}  , a_{A2}  , a_{B2} ]^T$:
\begin{equation}
\label{eq:H0_mangetic2}
\Hca_0 =
\begin{pmatrix}
V/2 & \gamma \sqrt{n} & \tp & 0 \\
\gamma \sqrt{n} & V/2 & 0 & 0 \\
\tp & 0 &  -V/2 & \gamma \sqrt{n+1} \\
0 & 0 & \gamma \sqrt{n+1} & -V/2
\end{pmatrix}.
\end{equation}
For $n \geq 1$ this leads to a spectrum that (as  function of $\gamma$) is very similar to the case without a magnetic field as a function of the absolute value of the momentum. The most important feature for us is that the gap is still present for these quantum numbers.

For $n=-1$ the spinor is simply $[0,0,0,|0\ra]$ which leads to a flat band (Landau level) at $-V/2$.
\begin{figure}[h]
\vspace{0.4cm}
\begin{center}
\includegraphics[width=0.8\columnwidth]{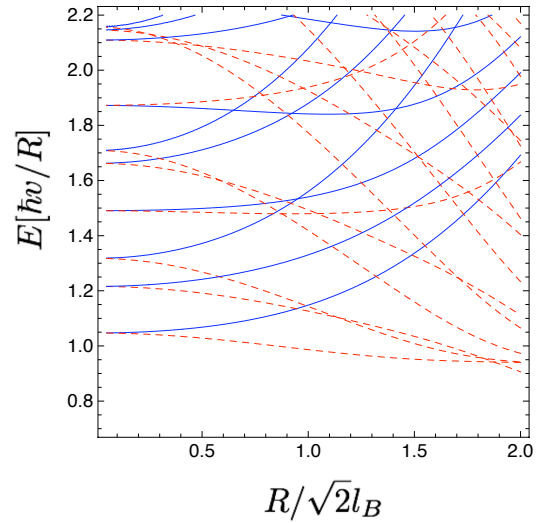}
\end{center}
\vspace{-0.4cm}
\caption{Energy levels in a relatively small bilayer QD (radius $R=25 \, \text{nm}$) as a function of the magnetic field. The other parameters are as follows: $t_{\perp}=0.4 \, \text{eV} =15.19 \hbar v/R$, $V=1.9 \hbar v/R$, $ U_0=1.52 \hbar v/R$ and $s=1$ (i.e. positive $B$-field). The solid and dashed lines are for different valleys.}
\label{fig:bilayerlevels1}
\end{figure}

The case $n=0$ is more interesting. In this case the spinor is of the form
$[a_{A1} |0\ra , 0 , a_{A2} |0\ra , a_{B2} |1\ra  ]$ and the resulting problem is the diagonalization of the matrix
\begin{equation}
\label{eq:H0_mangetic3}
\Hca_0 =
\begin{pmatrix}
V/2  & \tp & 0 \\
\tp &   -V/2 & \gamma  \\
0 &  \gamma  & -V/2
\end{pmatrix}.
\end{equation}
This leads to three levels as a function of $\gamma$. Two start out at $\pm \tp$ at $\gamma = 0$ and evolve smoothly over to $\pm \gamma$ as $\gamma \rightarrow \infty$. These two levels are therefore much like the case $n\geq1$. The \emph{most interesting level} starts out at $-V/2$ at $\gamma = 0$ and goes smoothly to $V/2$ as $\gamma \rightarrow \infty$. It is easy to see that the level crosses zero at $\gamma = \sqrt{\tp^2+ V^2/4}$. This Landau level that crosses the gap can also be seen in Fig.~3 of Ref.~\onlinecite{PPV07} and has important consequences for the levels in the QD, as we will discuss in the next subsection.

\subsection{Results for the bound state levels}
\label{rbsl}

The bilayer QD is in many ways similar to the single layer QD discussed above, but there are also important differences in the physics.

The most important result of our study can be seen in Fig.~\ref{fig:bilayerlevels1} where we display the energy levels of a dot as a function of the magnetic field. At zero magnetic field, the degeneracy of the levels in the two valleys is clearly displayed. With increasing the magnetic field, the orbital degeneracy is lifted. The symmetry of the levels is analogous to the case of the single layer discussed above. The states that are degenerate at zero field are related by time-reversal symmetry which means that they correspond to opposite values of angular momentum $\pm m$ in different valleys. The typical effective time-reversal symmetry of $\pm m$ within one valley is already broken by a ``mass'' term (which breaks the inversion symmetry of the bilayer) in a similar manner to the case of Neutrino billiards considered by Berry and Mondragon.\cite{Berry1987}
\begin{figure}[h]
\vspace{0.4cm}
\begin{center}
\includegraphics[width=0.8\columnwidth]{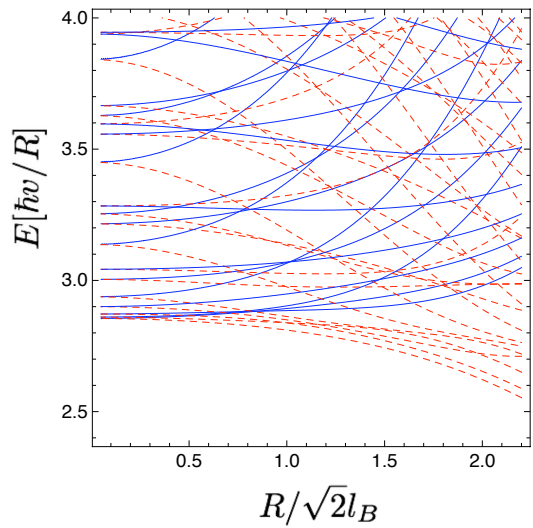}
\end{center}
\vspace{-0.4cm}
\caption{Energy levels in a relatively small bilayer QD with the same parameters as in Fig.~\ref{fig:bilayerlevels1} except that the bias field $V$ is about three times larger: $V = 6 \hbar v/R$.}
\label{fig:bilayerlevels2}
\end{figure}

\begin{figure}[h]
\vspace{0.4cm}
\begin{center}
\includegraphics[width=0.8\columnwidth]{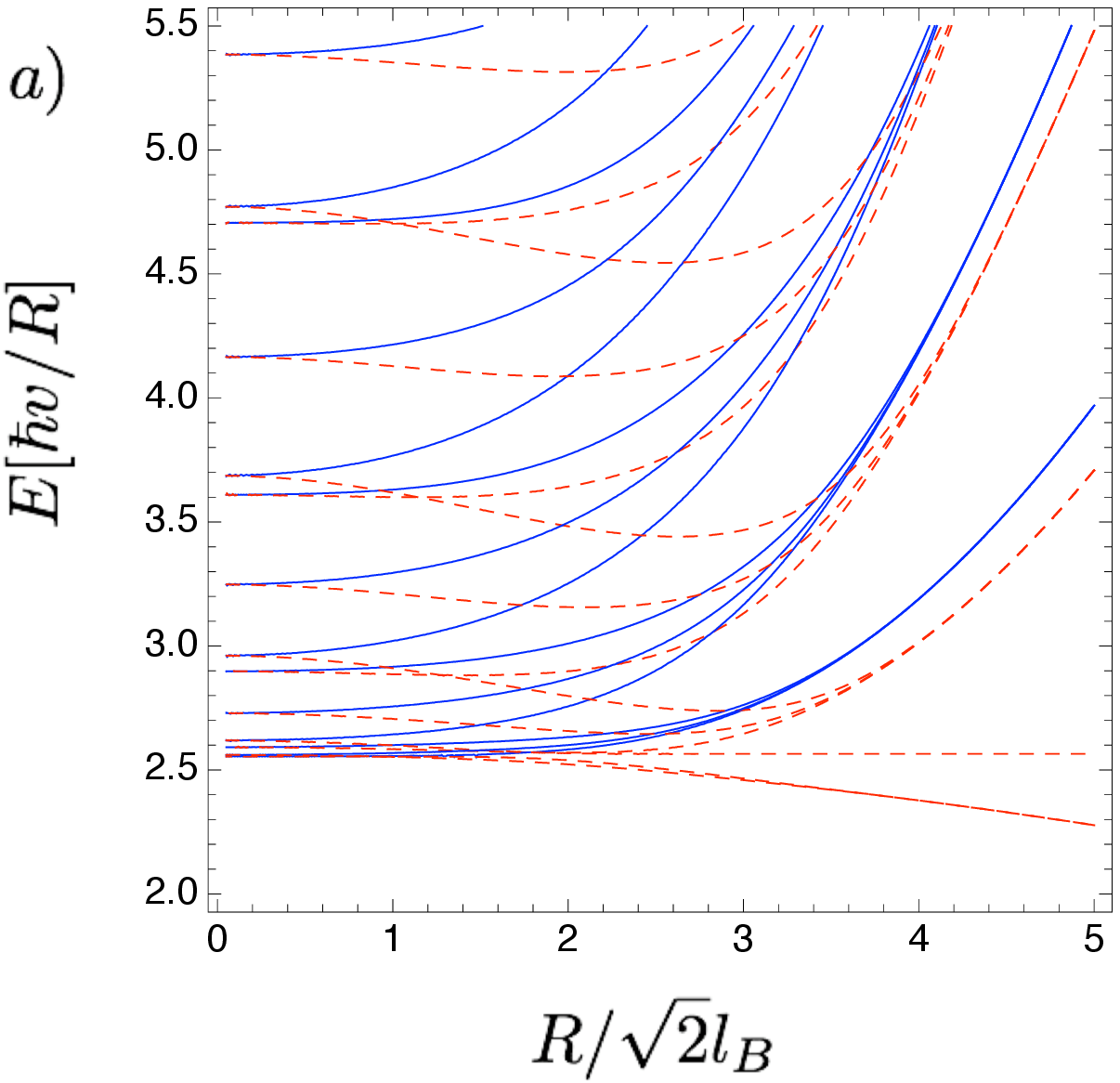}
\end{center}
\begin{center}
\includegraphics[width=0.8\columnwidth]{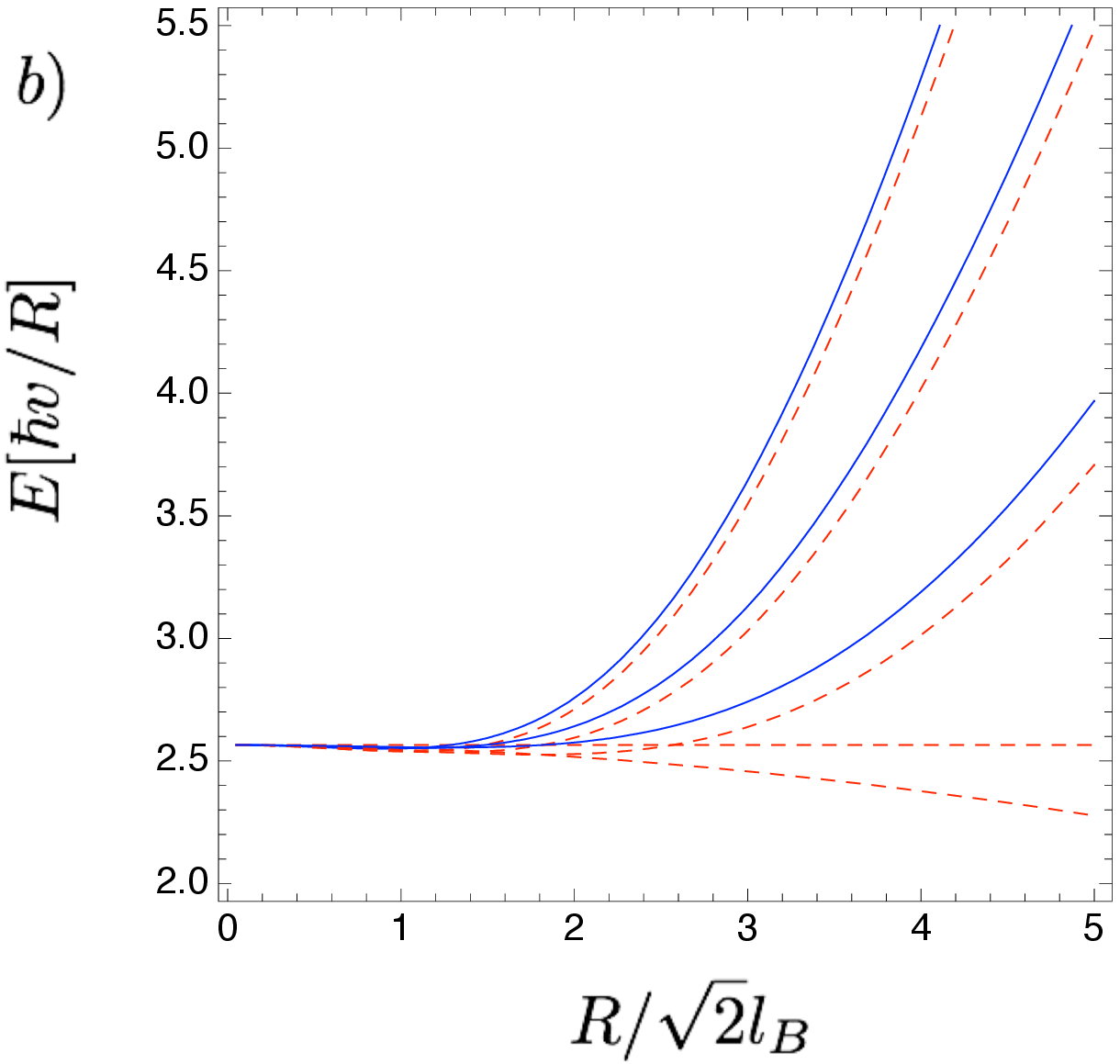}
\end{center}
\vspace{-0.4cm}
\caption{a): Merging of bilayer QD levels to the bulk LLs as function of magnetic field for a relatively large bilayer QD with R=67.48 nm and  $t_{\perp}=0.4 \, \text{eV} =41  \hbar v/R$, $U_{0}=3.5 \hbar v/R$, $V=5.13 \hbar v/R$ for $m=0,\pm 1$ and $s=1$ (i.e. positive $B$-field). Full lines are for $\tau=+1$ and dashed lines are for $\tau=-1$. b) Bulk Landau Levels (LL) which are approached almost perfectly at high fields in this parameter regime. Note that the $n=0$ LL (see subsection III B) crosses the gap with a negative slope, whereas the other LLs ($n=1,2,3$) have positive slope (blue and red denotes the two valleys). There is also a flat LL ($n=-1$) at $V/2$  similar to the single layer case. }
\label{fig:bilayer_LL}
\end{figure}

An important feature of the bilayer as opposed to the single layer is the unconventional Mexican hat-like dispersion relation near the band edge. This is most apparent for a large value of the bias field $V$. An example of the level structure for such a dot is shown in Fig.~\ref{fig:bilayerlevels2}. It is clear that there are many closely spaced levels near the band edge. This is a feature of the enhanced density of states near this particular energy.\cite{Nilsson2007} It is also crucial to note that the trigonal distortion term (which breaks the cylindrical symmetry and in principle couples all states with angular momenta $m$, $m \pm 3$, $m \pm 6$, $m \pm 9$,  $\ldots$) is a particularly relevant perturbation for the degenerate states close to the band edge. For states away from the band edge for which the energies of the coupled states are different in energy, the trigonal distortion term can be treated as a perturbation and we do not expect that the energy levels will be much affected. More explicitly, we use a cylindrically symmetric dispersion relation whereas the real dispersion relation including the trigonal distortion term, which is parametrized by $v_3$ ($v_3 \approx 0.1$ in graphite but it has not yet been measured in bilayer graphene), is not. From the expressions in Ref.~\onlinecite{Nilsson_long_2008} (zero magnetic field) we find that above the momentum scale $p_c \sim v_3 \tp$ the cylindrically symmetric term that we keep is dominating over the trigonal distortion term in the Hamiltonian and does hence provide a reasonable zeroth order approximation. It is not trivial to convert this momentum scale into an energy in general because of the Mexican hat structure. But for the parameters we use in Figures~\ref{fig:bilayerlevels1} , \ref{fig:bilayerlevels2}  and \ref{fig:bilayer_LL} the associated momentum is larger than $p_c$ for energies above the region of the Mexican hat (i.e. above $V/2$) where the energy becomes a monotonously growing function of momentum. Therefore,  we believe that our model captures the relevant physics above the Mexican hat. We note that at finite magnetic fields, it is known that the trigonal distortion quickly becomes less important with growing magnetic field in an unbiased bilayer. \cite{Falko2006a} We therefore expect that for the large field regime, the corrections are small at all energies.
Additional subleading parameters (such as $\gamma_4$ which introduce an electron-hole asymmetry into the spectrum \cite{Zhang_2008_bilayer}) will also shift the level positions slightly. 

For a large QD, it is also possible to reach the regime where the dot levels are described by the Landau levels. This feature is seen in Fig.~\ref{fig:bilayer_LL}a) where we display the bound states for $m=0,m=\pm 1$ for large magnetic fields. Note that the QD levels tend to approach the bulk Landau levels displayed in Fig.~\ref{fig:bilayer_LL}b).
In a smaller QD it is hard to reach the Landau level limit for moderate magnetic fields.

\begin{figure}[h]
\vspace{0.4cm}
\begin{center}
\includegraphics[width=0.8\columnwidth]{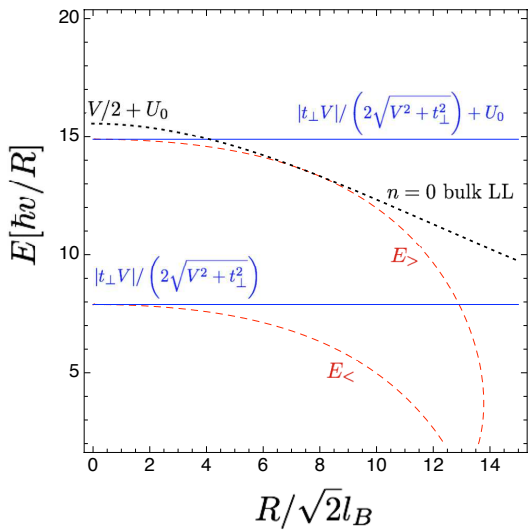}
\end{center}
\vspace{-0.4cm}
\caption{Effective bandwidth of the bilayer QD defined by the area between the lines $E_{>}$ and $E_{<}$ (dashed lines) as a function of magnetic field for $V=17.1 \hbar v/R$ and $U_{0}=7 \hbar v/R$, $s=1$, $t_{\perp}=41 \hbar v/R$ and for valley $\tau= -1$. The dotted line displays the $n=0$ bulk LL (in valley $\tau=-1$) that crosses the QD bandwidth corresponding to the band gap at zero magnetic field (full lines). This bulk LL presents an escape channel for the bilayer QD if bound states cross this bulk LL. Note that no bulk states overlap with the effective bandwidth (from the same valley).}
\label{fig:bilayerwindow}
\end{figure}

Another important feature of the bilayer for designing a QD is the existence of an anomalous LL that crosses the gap, see subsection III.B and Fig.~\ref{fig:bilayer_LL}b). The character of bound states changes, when the square root of $\kappa_{<,>}^2$  in Eq.~(\ref{kappas}) changes sign, which occurs at energies
\begin{equation}
\label{kappa2a}
E_{<}= \frac{s\tau V}{l_{B}^2(\tp^2+V^2)}\pm\sqrt{\frac{V^2\tp^2}{4(\tp^2+V^2)}-\frac{\tp^2}{l_{B}^{4}(\tp^{2}+V^2)^2}},
\end{equation}
and at $E_{>}=E_{<}+U_{0}$. These lines are shown in Fig.~\ref{fig:bilayerwindow} as a function of magnetic field. The area between $E_{<}$ and $E_{>}$ define an effective (valley-dependent) bandwidth for the QD. Indeed, at $B=0$, we obtain only bound states for energies above $|t_{\perp} V|/2\sqrt{t_{\perp}^2+V^2}$ and below  $|t_{\perp} V|/2\sqrt{t_{\perp}^2+V^2}+U_{0}$ which correspond to the conduction band minimas inside and outside the QD, respectively. Within this bandwidth, the two $\kappa^2$'s inside the QD ($\kappa_{<}$) are purely real and the two $\kappa^2$'s outside the QD ($\kappa_{>}$) have an imaginary part. The physical meaning of $\kappa/(\sqrt{2}l_B)$ is most transparent in the limit of zero magnetic field where it becomes the inverse decay length of the wave function.\cite{note1} Thus states within the bandwidth correspond to a decaying wave outside that is matched to one propagating and one decaying wave inside of the dot (this is true for energies such that only one band is allowed inside of the dot whereas the other band requires the momentum to be imaginary).

The bound state energy window at $B=0$ is crossed by the $n=0$ bulk LL, as shown in Fig.~\ref{fig:bilayerwindow}. When QD bound states cross this bulk LL, the QD becomes "leaky" and electrons can escape into the bulk. The effective bandwidth defined by the area between the lines $E_{>}$ and $E_{<}$ is never crossed by a bulk LL and defines therefore a "safe" zone for QD bound states.
Note, however, that bound states do exist also outside of this effective bandwidth at finite magnetic field, since the the bulk spectrum (LLs) becomes discrete. This means that evanescent waves continue to exist in the "bulk" region when they are not degenerate with a bulk LL (much like the edge states that are present between the LLs in the integer Quantum Hall Effect). However, this regime is not ideal for QDs, due to the leakage via nearby bulk states as described above.

We point out that QDs in bilayer graphene in connection with a magnetic field again allow for a controlled  tuning of level degeneracies. The values used in our plots correspond to realistic values of the gap voltage $V$ and the inter-layer coupling $\tp$. \cite{Castro_PRL_2007} Besides similarities to the single layer case studied in section II, the bilayer QD shows very interesting additional features. The size of the gap $V$ can influence the size of the level spacing drastically when the energies are close to the band edge, where a Mexican-hat like structure is formed which is more pronounced at larger $V$.

\section{Applications of valley splittings}

In this section we discuss the implications and their use of the broken valley degeneracy by a magnetic field in gate-tunable graphene QDs.

\subsection{Consequences for valleytronics}

The valley index $\tau=\pm 1$
 can be thought of as eigenvalues of the operator ${\bm\nu}\cdot{\bm \tau}$ where ${\bm \nu}$ is a unit vector on the Bloch sphere and ${\bm \tau}$ the vector of Pauli matrices. \cite{Carlographenereview}
 The operator ${\bm \tau}$ is called the valley {\it isospin}. If the two valleys are uncoupled, we have ${\bm \nu}=\hat{z}$.
It has been pointed out that the valley isospin ${\bm \tau}$ could be used as a controllable degree of freedom like the electron spin ${\bm S}$ is used in spintronics applications which coined the name valleytronics. \cite{Rycerz2007}
The main motivation to use the valley degree of freedom as a new unit of information in graphene instead of the sublattice pseudospin ${\bm \sigma}$, is the fact that the valley degree of freedom is preserved in the absence of short range scatterers (whereas ${\bm \sigma}$ is not), provided e.g. by the graphene edges.
However, the manipulation of the valley isospin is not as straightforward as for the real electron spin since the valley isospin does not directly couple to a magnetic field as does the real spin via the Zeeman interaction. However, since the valleys are related by time-reversal symmetry, the valley degeneracy can also be broken in principle by applying a magnetic field which we have shown in this work. However, a magnetic field alone is not enough since it breaks only degeneracies within {\it different} valleys. The so called {\it effective} time reversal symmetry ${\bm p}\rightarrow -{\bm p}$ and ${\bm \sigma}\rightarrow -{\bm \sigma}$ within each valley should also be broken. This is achieved by quantum confinement induced by a boundary that does not couple the valleys, \cite{Recher2007} which is the case for the gate-tunable QDs proposed here (see also Sec.~II B).

In this work we have shown that the valley degeneracy, and more generally, the orbital degeneracy is controllably and efficiently broken by a magnetic field. In the case of the valley splitting $\Delta_{K,K'}$, we can take advantage of the anomalous LLs which are approached by the QD states at large $R/l_{B}$. As we have discussed in sections II and III, there exists a flat LL at the gap value that is only present in one of the valleys (its other valley partner is at negative energy, i.e. in the valence band). In the bilayer QD, we have in addition a state that crosses the gap with increasing magnetic field.
We can estimate typical values for $\Delta_{K,K'}$ by comparison with our plots for QD bound states. In the single layer, we obtain from Fig.~\ref{singlelayerfig}b) for a dot radius of $R=67.48$ nm and for $B\sim 4.6$ T, a valley splitting $\Delta_{K,K'}$ of about $24.4$ meV between the valley-polarized ground state and the first excited states (from the other valley).  For the same QD size and for the same $B$-field, we obtain for the bilayer QD shown in Fig.~\ref{fig:bilayer_LL}a), a splitting between the approached $n=0$ LL and the $n=1$ LL from the other valley of $\Delta_{K,K}\sim 8$ meV. Both values for $\Delta_{K,K'}$ are much larger than temperature and can be probed in tunneling transport through QDs.

Such QDs could be used as very efficient {\it valley-filters} in transport through the QD. In contrast to earlier proposals of valley filters in zigzag ribbons in single layer graphene, \cite{Rycerz2007} and topologically confined states in bilayer graphene \cite{Martin2008} without magnetic field, the present setups would work as filters that break time-reversal symmetry and therefore function also in closed QD systems where Coulomb blockade (CB) effects can be used to operate at the single (valley-)spin level much like for ordinary spin-filters in QDs. \cite{Recher2000} CB effects (and therefore single electron tunneling) become prominent if the charging energy exceeds the temperature and for weak coupling to the leads (tunnel resistance $\gg h/e^2$). We estimate the charging energy $E_{C}$ as a function of the QD radius $R$ as $E_{C}=e^2/C$ with capacitance $C=8\varepsilon_{0}\varepsilon_{{\rm eff}}R$ where $\varepsilon_{{\rm eff}}=(1+4)/2$, including the dielectric constant for ${\rm Si}{\rm O_{2}}$ and vacuum. \cite{Stampfer2008b} For $R=67.48$ nm, this gives $E_{C}\sim 12$ meV. The tunneling rate in and out of the QDs could be tuned by gates. We note that two valley-filters in series can be used as a {\it valley valve}, \cite{Rycerz2007}
where the valley polarization of one of the QDs should be reversible. This can be achieved by either reversing the sign of the magnetic field, or more easily, by gates such that one QD can be tuned from a hole-doped QD to a n-doped QD (and vice versa). In this way, the valley isospin of QD states at resonance with the the leads can be changed in one QD, thereby probing the polarization of the other QD. We note that the {\it presence} of such a valley splitting could be probed by electron transport since the level degeneracy is changed with increasing magnetic field. \cite{Recher2007} Since the valley splitting $\Delta_{K,K'}$ acts like a Zeeman field for the valley isospin, experiments that measured the relaxation time $T_1$, \cite{Hanson2003} and the read-out of a single electron spin \cite{ElzermanNature} in GaAs QDs could be performed in a similar way in gate-tunable graphene QDs in order to measure the valley relaxation time and valley polarization in such QDs (besides the detection measurements for real spin).

\subsection{Consequences for spin qubits}\label{spinqubits}

The use of the spin 1/2 degree of freedom of single electrons as qubits
\cite{LossDiVincenzo1998} is usually combined with a proposed coupling
of adjacent localized spin qubits via the Heisenberg exchange
interaction. \cite{BLD1999}  Carbon may offer
relatively long spin coherence times due to the sparseness of nuclear
spins and potentially also due to the weakness of the spin-orbit
coupling in these materials. \cite{Huertas2006,Min2006}
However, spin qubits in graphene QDs, \cite{Trauzettel2007}
need to deal with the valley degeneracy in these materials which
can interfere with exchange coupling.  This can be understood using
the following simple model for two electrons in adjacent graphene QDs.
Suppose that the valley degree of
freedom, unlike the spin in this case, is not well under control,
and, because it is degenerate, each electron is in an incoherent mixture
of the two valley states K and K', with equal probability,
$\rho_\tau=(|K\rangle\langle K|+|K'\rangle\langle K'|)/2$.
The density matrix of the two electrons in adjacent QDs can thus
be written as
$\rho_\tau=(|KK\rangle\langle KK|+|K'K\rangle\langle
K'K|+|KK'\rangle\langle KK'|+|K'K'\rangle\langle K'K'|)/4$.
Including spin, the density matrix is then
$\rho=\rho_\tau\otimes |\varphi\rangle\langle \varphi|$
where $|\varphi\rangle$ is an arbitrary pure two-spin state.
At this point, the spin and valley degrees of freedom are
uncoupled, and while the spin can maintain its coherence, the
valley isospin may at the same time be in an entirely incoherent state.
The problem arises because there will be a tunnel-coupling
mediated exchange coupling $J\neq 0$ if the electrons are both in the same
valley $|KK\rangle$ or $|K'K'\rangle$ but there will be no such
coupling ($J=0$) in the cases where the electrons are in different valleys,
i.e. $|KK'\rangle$ and $|K'K\rangle$.  The reason for this is that
the exchange coupling relies on the Pauli exclusion principle
which only matters in case that both electrons can occupy the same
orbital.  Here, we assume that the inter-dot tunneling conserves
the valley isospin (however, we note that a similar conclusion would be obtained in the non-conserving case). Suppose we apply the exchange
coupling such that it generates a SWAP operation that exchanges the
two spin qubit states. \cite{LossDiVincenzo1998}  This SWAP operation
will be conditional on the valley state.  Assuming an initial spin
state $|\varphi\rangle = |+-\rangle$, where
$|\pm\rangle = (|\uparrow\rangle \pm |\downarrow\rangle)/\sqrt{2}$,
we find, after the SWAP, the state
$\rho'=(|KK\rangle\langle KK|+|K'K'\rangle\langle K'K'|)/2\otimes
|-+\rangle
+(|KK'\rangle\langle KK'|+|K'K\rangle\langle K'K|)/2\otimes
|+-\rangle$.
If the valley degree is traced out, we find that
$\rho'=(|-+\rangle \langle -+|  +  |+-\rangle \langle +-|)/2$.
With this, the phase coherence of both spins decays
$\langle\sigma_x^i\rangle = 0$, $i=1,2$, due to the coupling to the
incoherent valley degree of freedom.  Even if the valley
isospin is coherent, a valley degeneracy will still lead to
spin-isospin entanglement, which for some purposes may be
interesting, but which essentially reduces the spin coherence
to some charge (valley) coherence time which can be expected
to be shorter.  If the valley degeneracy can be lifted, as
proposed here, one can avoid this entanglement and possible
spin decoherence processes that arise from it.
We note that orbital degeneracies in the same valley lead to similar problems for the exchange
coupling of neighboring spins. We therefore should operate at moderate magnetic fields such that the LL regime is
not reached.

\section{Conclusions}

We have studied the bound states of QDs in gapped single- and bilayer graphene in the presence of a homogeneous magnetic field. Due to the absence of sharp graphene edges, the valleys are well defined in these QDs.  We have shown that these realistic structures would allow us to control the valley degeneracy by the magnetic field. This has important consequences for spin or valley-quantum computing, where the breaking of orbital (or valley) degeneracy is absolutely crucial. Besides similarities between the two systems, we also found crucial differences that can be traced back to an anomalous Landau level (LL) in the gapped bilayer that crosses the gap and which can provide an escape channel for QD bound states into the bulk at large magnetic fields if they cross this LL. In addition, the level spacing size close to the band edge, crucially depends on the strength of the applied voltage in the bilayer QD, which is due to a "Mexican hat" form of the bulk bandstructure. These features have important implications for finding the ideal parameter range for useful QDs. We also discussed possible applications of such QDs for the emerging fields of valleytronics and spin qubits in graphene.

\acknowledgments
We thank C. W. J. Beenakker for helpful comments and discussions.
P. R. and J. N. are supported by the Dutch Science Foundation NWO/FOM. G. B. is supported by the Swiss National Science
Foundation (SNF) via grant no. PP002-106310 and by the German Research Foundation (DFG)
via Forschergruppe FOR 912 ("Coherence and relaxation properties of electron spins "). B. T. is supported by the German Research Foundation (DFG) via grant no. Tr950/1-1.

\bibliography{graphite}

\end{document}